Article

# Towards a GDPR-Compliant Blockchain-Based COVID Vaccination Passport

AKM Bahalul Haque *, Bilal Naqvi , A. K. M. Najmul Islam and Sami Hyrynsalmi

Software Engineering, LENS, LUT University, 53850 Lappeenranta, Finland; syed.naqvi@lut.fi (B.N.); najmul.islam@lut.fi (A.K.M.N.I.); sami.hyrynsalmi@lut.fi (S.H.)
* Correspondence: bahalul.haque@lut.fi

**Abstract:** The COVID-19 pandemic has shaken the world and limited work/personal life activities. Besides the loss of human lives and agony faced by humankind, the pandemic has badly hit differ before departure and on arrival, and voluntary quarantine, were enforced to limit the risk of trans- mission. However, the hope for returning to a normal (pre-COVID) routine relies on the success of the current COVID vaccination drives administered by different countries. To open for tourism and other necessary travel, a need is realized for a universally accessible proof of COVID vaccination, allowing travelers to cross the borders without any hindrance. This paper presents an architectural framework for a GDPR-compliant blockchain-based COVID vaccination passport (VacciFi), whilst considering the relevant developments, especially in the European Union region.

## 1. Introduction

Humankind is facing the challenges and agonies posed by the coronavirus disease (COVID-19) pandemic for more than a year [1]. Since its beginning, more than 130 million people have contracted this disease worldwide, and the number continues to grow daily [2]. The number of people affected by COVID-19, whether in the form of losing their loved ones, losing their jobs, abandoning their businesses, is far greater than 130 million [3]. Among the many industries affected by COVID-19, the travel and tourism industry got the biggest hit. Due to the pandemic nature of the disease, affected countries had to impose lockdown procedures to stop the spread of the virus [4]. Work has mostly shifted to remote means wherever possible, and the only ray of hope towards returning to normal life lies in the success of the vaccination drives. However, the vaccination of every individual in the world is going to take time [5,6].

To open the regions and to partially allow various sectors to revive, a need for vac- cination passports was realized. It is worthwhile to mention what we mean by a vaccine passport; it is a printed or digital certificate that can certify that a person is unlikely to either catch or spread disease. From the COVID-19 perspective, the proposed passport would certify that the holder has been vaccinated. However, such an implementation could also be used for certifying vaccination against other diseases such as polio, tuberculosis, measles, etc.

Considering the COVID-19 perspective, the passport could allow governments to relax restrictions, allowing vaccinated people to travel in planes, attend concerts and sports events, go to work, or dine out, etc. Several governments in the world including in Asia, Europe, and the Middle East have plans of issuing such passports to vaccinated people to allow them to travel if desired without having to go through COVID-19 related traveling protocols. Specifically, in the EU region, the need for vaccine passports was realized from tourism-dependent southern countries. The EU commission has plans of having one such passport before the summer of 2021 to partially open the region for tourism [7,8].

When considering the development of such a passport, the requirements such as integrity, verifiability, traceability, and trust by design are considered as essentials. With such requirements, there are two technologies as candidates for consideration, (1) public key infrastructure (PKI), and (2) blockchain. Each of these has its advantages and limitations. Using PKI would have meant that each health authority issues a vaccination certificate signed with the private key and verifiable using a public key at the border controls. The limitations of such a solution include (1) issuance of a new certificate after administering every dose, (2) lack of traceability by design, and (3) issuance of a new certificate in case of loss/deletion. Moreover, with PKI, a separate mechanism for the storage of the vaccination records by health authorities would be required, as the digital signatures hold good for authenticity and integrity purposes, and do not promise the ability to have tamper-proof records and transactions. However, blockchain aims to deliver these aspects by addressing trust issues while also enabling data integrity, availability, and verifiability [9]. Blockchain uses the concept of decentralized architecture to build an interlinked chain of blocks. The block data is tamper-proof and can only be deleted by destroying the whole chain. Data inserted in the block is verified by the blockchain network's participating authorities using specific consensus algorithms [10]. The use of blockchain has been also suggested for solving a variety of issues involving the healthcare sector [11], therefore, we considered blockchain for the development of COVID vaccination passports. With blockchain, issues such as the issuance of a new certificate after administering each dose, losing the certificate are not applicable since the data held on the chain itself is updated and therefore, can be verified. Moreover, the use of blockchain also enables a single mechanism for verification and storage of the vaccination record. Furthermore, blockchain technology is used in different sectors including, identity management [12], smart homes [13], agriculture [14], power management [15], access control [16], and others.

A smart contract is another important element that can be applied with blockchain for verifiability, traceability, authentication, and ensuring trust. Smart contracts enable the automatic execution of instruction based on the fulfillment of certain conditions [17]. In this work, we are going to leverage both blockchain and smart contracts for presenting an architectural framework for COVID vaccination passports. This vaccination passport will enable the concerned authority to verify the vaccination information of an individual anytime and anywhere. From the EU's perspective, recommendations from General Data Protection Regulations (GDPR) have been taken into consideration while formulating the architectural framework for the COVID vaccination passport. We also referred to the existing literature and conducted interviews with the personnel from Finland's blockchain industry for identification of the relevant protocols and mechanisms for the design of GDPR-compliant blockchain solutions [18,19].

The remainder of the paper is organized as follows: Section 2 presents the background about blockchain technology, smart contracts, and GDPR; Section 3 outlines methodology for formulation of the proposed architectural framework; Section 4 presents the architec- tural framework for COVID vaccination passport and summarizes how it tackles the GDPR compliance issues related to the proposed architecture; Section 5 presents the research contributions and limitations, and Section 6 concludes the paper.

## 2. Background

In this section, we discuss some of the aspects related to blockchain technology imperative to be discussed before presenting the proposal for a vaccine passport. As stated earlier, we utilize smart contracts for the architectural framework. Therefore, we also discuss some details about smart contracts.

*2.1. Blockchain*

Blockchain is very similar to a database or ledger. However, it is more advanced and inherits the elements of security due to its cryptographic nature [20]. A decentralized network is a collection of many centralized networks interconnected in a distributed manner. All the nodes of the network share their data and make a copy of the whole ledger. Blockchain is an ever-growing, interlinked chain of blocks that stores the records of all the transactions committed. Each block is linked to the next with the help of a cryptographic hash value of the previous block. Besides hash value, the blocks also contain transaction data such as the timestamp. Blockchain makes it possible to make transactions without any intermediaries, thus making it indeed a decentralized technology. It also uses public- key or asymmetric cryptography and consensus algorithms to ensure user security and immutability [21]. In Blockchain, all the nodes act as peers and they are treated equally. The nodes can share data directly and do not need any centralized server to be able to send or receive.

Blockchain is a reliable space to store and share data in a transparent manner. The path of the transaction can be checked or verified by the other nodes in the network. If any of the nodes think a transaction is risky, the node can cancel the transaction instantly [22]. There are mainly three types of blockchain, (1) permissionless blockchain, (2) private or permissioned blockchain, and (3) consortium blockchain. It is relevant to state that the architectural framework presented in this paper utilizes permissioned blockchain.

(1) Public Blockchain: A public blockchain is a non-restrictive, permission-less distributed ledger-based system. Anybody having access to the Internet can register and sign in to the blockchain platform. Each member can check the transaction and verify it. Also, anyone can participate in the process of getting consensus. Bitcoin and Ethereum are the relevant examples [6]. A public blockchain also demands a significant amount of computing capacity, because maintaining a distributed ledger on a wide scale takes a significant amount of processing power. Each node in the network needs to solve a complicated cryptographic puzzle to verify that all nodes in the blockchain have coordinated to reach a consensus [7].

(2) Private Blockchain: It is restrictive in nature and requires permission, which limits it to work only for private networks. It has strict authorization management for accessing data. Only the selected participant has access to the system within an enterprise. Security, authorization, and accessibility are controlled by enterprises unlike public blockchain [9]. Private blockchains are easier to scale up, cut down costs, and feature greater transactions. This approach can enable the organization to have more control over the network and establish integrity by removing anonymity.

(3) Consortium Blockchain: This is another type of permissioned blockchain, which is also referred to as a semi-decentralized blockchain. In this type of blockchain, predefined approved parties (nodes) participate in the block validation process. If a group of enterprises wants to implement a blockchain network and requires only a few approved parties to participate in the process, this type of blockchain plat- form can be implemented. Consortium blockchain allows the implementation of customized organizational policies. In addition to these characteristics, depending on the requirements of customers (organizations), services of the blockchain (i.e., inquiries about the information in the blockchain) can be open to the public or other authorities who are not part of the blockchain. The proposed architecture is based on elements of consortium blockchain enabling the concerned authorities to perform the needful tasks and inquire about any information residing on the blockchain. More- over, the network can be easily expanded and if required, the participating authorities can be added [23–25]. The factors which further motivate our choice of consortium blockchain are as follows:

- It is a permissioned blockchain, therefore, only authorized nodes can validate the blocks.
- It is highly scalable allowing any party to get involved.
- Though it is a permissioned blockchain, its services can be seen by the public by using customized APIs.
- It allows spreading across a specific area or geographical region, allowing control over data transfer and processing.

- Potential data breaches can also be investigated easily since the participating authorities are fixed and finite.

### 2.2. Smart Contracts

A smart contract is an instruction set that automatically executes once certain user-defined conditions are fulfilled. Smart contracts have the benefit of allowing two distrusting parties to proceed with a transaction, without the need for a trusted third party such as a lawyer or bank [26]. Fundamentally, the smart contract is a programmed protocol that executes based on contract conditions and agreements. If a smart contract is used in a blockchain, it becomes immutable. For this reason, blockchain and smart contracts could be the right combination for ensuring trustworthiness and transparency since the use of blockchain can make smart contracts immutable as well as visible to the relevant stakeholders. In addition, the concept also preserves the logical connections between contractual clauses through the logical flows in the program. Smart contract's life cycles can be separated into four parts–creation, deployment, execution, and completion. This cycle continues consecutively. Smart contracts can guarantee proper access control and contract enforcement. The intended statement will automatically execute only when any condition of the smart contract is satisfied. An automatic penalty system might also be there if anyone breaches the contract [27,28].

The proposed architecture is based on smart contracts along with the consortium blockchain. A smart contract can also be used for consent collection [10]. Since the proposed architecture uses the hash of the personal data, the same can be used in the smart contract for consent collection purposes. The identifiable data is stored in the off-chain (traditional storage) so that there will be no issues in deleting that data on request. No personally identifiable information will be stored on the blockchain. Instead, a hash of some identifiers will be stored, which does not reveal any information about the data subject.

Apart from the above-mentioned aspects, the smart contract provides the follow-ing benefits.

- A smart contract promotes traceability when it is used to collect the status of the information on the blockchain. The transaction log of the smart contract can also be used as a travel log to trace and limit potential COVID exposures and infections.
- The smart contracts execute automatically and the traveler will not have to provide any extra details for that.
- Verification with the help of a smart contract is faster than manual verification since it does not require human intervention. Moreover, there are no intermediaries involved.

### 2.3. GDPR

GDPR (General Data Protection Regulation) is designed to protect the rights and freedom of European Union (EU) citizens. GDPR enables the protection and privacy of the personal data associated with EU citizens. The regulation requires all the concerned parties dealing with the personal data of EU citizens to comply with the GDPR [29]. The primary policy of GDPR is, the companies must inform the data subjects in an understandable manner about the data collection and usage procedures. The data subject possesses the ultimate power to give consent about the data collection and deletion. However in certain circumstances (falling under the purview of Article 6.1.e) the right to deletion does not rest with the data subject, but the authority lies with the data controller

The rapid growth of information and communication technologies is giving rise to the volume of personal data day by day. Healthcare, educational institutions, supply chain industry, and the agricultural sectors are becoming digitalized with time and hence data protection and privacy issues are going to be a critical concern [30,31].

The proposed vaccination passport is intended for the EU region, therefore, GDPR compliance issues are imperative to be considered [32,33]. Among all the compliance issues, data deletion, modification, defining the controller and processor are the most crucial ones. Moreover, we also analyzed existing literature that closely relates to the GDPR compliance issues; to mention, Al-Zaben et al. [34] proposed an off-chain mechanism for storing original data instead of storing it in the blockchain. Tatar et al. [35] and Bayle et al. [36] recommended a similar approach for storing personal data off-chain for compliance with GDPR.

Moreover, with blockchain being a decentralized network, identification of the data processor and controller is a crucial problem. GDPR explicitly identifies the data controller and processor because it creates transparency. Buocz et al. [37] recommend miners as a data controller, but only the miners with significant financial leverage over the network can be the controller. The financial leverage can help with enforcing policies if needed throughout the network. Additionally, Dutta et al. [38] and Kondova et al. [39] proposed the use of a joint controller. Dutta et al. [38] also proposed using miners as data processors in case of permissioned blockchain. Both the recommendations closely align with the proposed scheme.

## 3. Methodology

### 3.1. Identification of Design Requirements

To formulate the architectural framework for the COVID vaccination passport, the first step was to identify the aspects that could have an impact on the design of COVID vaccination passports. To identify the design requirements, we collected the requirements from different sources and activities.

(1) We analyzed the two existing blockchain solutions deployed for verification and integrity checks using the case study research method [40]. We collected technical and marketing documentation of these blockchain solutions. One of the solutions is based on multi-tenant blockchain-based proof of content integrity by X Group. The salient feature of the proposed solution is trust, and it focuses on how to prove the stored information, or its metadata has not been modified. The solution is based on storing proof of integrity onto the blockchain. The solution is very much relevant to the context considered in this paper. However, the other solution which was analyzed is a blockchain-based platform for exchanging business documents in a machine-readable format. All transactions are validated to ensure high quality of data. In a broader perspective, the solution assists in managing the supply chain while maintaining interoperability between different parties. The aspect of GDPR compliance was very important and how these issues have been tackled in the solution was of prime significance.

(2) We collected and analyzed data from expert interviews working in six leading blockchain companies operating in Finland. The data were collected using semi-structured interviews. The average length of the interviews was approximately 1 h. To identify how practitioners tackle the GDPR challenges in their use cases, the data from interviews was content analyzed utilizing qualitative coding [41]. Coding is a technique employed in qualitative research for analyzing data items to identify concepts and find a relation between the data items. There are different approaches to qualitative coding, however, the one used in this case is data-driven coding or open coding.

(3) We have analyzed the data collected from 5 incoming travelers in Finland during this pandemic time. The travelers are from outside the European Union. They explained their experiences at the Helsinki Vantaa Airport and immigration. One of the travelers mentioned: "When I arrived at Helsinki Airport, I had to show my COVID negative certificate to the health authority at the airport in the arrival section. They were stationed with desks and chairs and tables. The health checking and verification is maintained by Finnish Institute for Health and Welfare. The booth was located was before the immigration departure. I had to show my identity documents as well as my COVID test report. After verifying my report with the QR code attached in my COVID report, I was instructed to do another COVID test after 72 h. If the result came negative, I could end my quarantine. After this station, there was no other verification at the immigration departure".

Another traveler who arrived a week later also experienced a similar situation. In addition, there is also COVID test-taking facilities at the airport, where one can get tested for COVID before entering Finland. One of the interviewed travelers did the COVID test at the airport since her COVID report was a little over 72 h. The Finnish health services inform each traveler that if anyone does not want to get tested for COVID (second time) after entering Finland, a 14-day quarantine is mandatory. If there is no sign of COVID infection, the quarantine can end. Interview data analysis (using open/inducting coding) epitomizes a few common requirements, for example, having a COVID certificate or previous COVID history, timestamp needed for the time of test along with a verification mechanism such as QR code embedded in the report.

(4) We studied and identified the relevant GDPR articles that have an impact on blockchain-based digital services.

(5) We have also analyzed publicly available records as part of the qualitative research methodology for identifying the design requirements [42]. The sources of public records are the border control of Finland [43,44], Sweden [45], and Norway [46]. In addition, we studied the updates on COVID-specific travel requirements on the EU Commission's and International Air Transport Association (IATA) [47] and FinnAir [48] websites. More than one source was used to remove any kind of bias that might affect the requirements. We extracted the key requirements for travel guidelines from these sources and considered those for designing the proposed architecture.

The key design requirements considered while formulating the architectural frame- work (identified after analyzing the data) are summarized as follows:

1. The citizens must be registered with local health authorities to get vaccinated. 2. The vaccinations are administered in specific hospitals under specific protocols. 3. Timestamps would be required to record the date of vaccination to ensure validity.

4. The qualitative analysis of the interviews data and existing blockchain solutions revealed the following:
   i. Facilitation of information flow between different authorities would be required. Different authorities involved in this could include immigration authorities, healthcare authorities, etc. The interview data highlighting this aspect was labeled and the code assigned was 'information flow'.
   ii. There is a need for increased trust so that the authorities use the service *(code: trust)*.
   iii. Consideration of security mechanisms in cases where it involves storing person- ally identifiable information (code: security mechanisms).
   iv. A procedure for verification that involves less computation (code: verifiability).

5. Complying with the GDPR articles (details discussed in the following sub-section).

### 3.2. GDPR Requirements

For the system to be used for collecting and processing data of EU citizens, it needs to be GDPR-compliant. Based on an analysis of the GDPR and blockchain, the articles that are discussed widely for compliance issues in terms of blockchain-based solutions are presented in Table 1 [49]. We identified the following three design principles for GDPR compliance.

1. Access to data should be traced.
2. Data subjects' consent should be collected using a smart contract.
3. Blockchain data needs to be deleted or modified upon request.
4. Data controllers and processors must be identified.

Next, we describe our proposed vaccination passport architecture that utilizes these identified design principles.

**Table 1.** Relevant General Data Protection Regulations (GDPR) articles for blockchain-based solutions.

| Article No. | Description |
|---|---|
| Article 5 | This article states the "Principles relating to processing of personal data". Some of the principles are "lawfulness, fairness and transparency", "purpose limitation", "data minimization", "accuracy", " integrity and confidentiality" etc. |
| Article 7 | The article demonstrates the data subjects' consent collection and management related to data processing behaviour and techniques. The beginning of the article in the regulation is stated as "Where processing is based on consent, the controller shall be able to demonstrate that the data subject has consented to processing of his or her personal data." |
| Article 9 | The article demonstrates "processing of personal data revealing racial or ethnic origin, political opinions, religious or philosophical beliefs, or trade union membership, and the processing of genetic data, biometric data for the purpose of uniquely identifying a natural person, data concerning health or data concerning a natural person's sex life or sexual orientation shall be prohibited" |
| Article 16 | "The data subject shall have the right to obtain from the controller without undue delay the rectification of inaccurate personal data concerning him or her" In addition, the data subject's right includes providing supplementary resources if it is needed based on the processing purpose. |
| Article 17 | "The data subject shall have the right to obtain from the controller the erasure of personal data concerning him or her without undue delay and the controller shall have the obligation to erase personal data without undue delay" Data deletion causes can be consent withdrawal, no legitimate grounds for data retention, data storage not needed. Moreover, the data controller also has to make sure of the appropriate technical measures for deleting the personal data before displaying it. |
| Article 18 | "The data subject shall have the right to obtain from the controller restriction of processing" under a few scenarios, for example, unlawful and illegitimate data processing, data storing is no longer required, etc. |
| Article 24 | "Taking into account the nature, scope, context, and purposes of processing as well as the risks of varying likelihood and severity for the rights and freedoms of natural persons, the controller shall implement appropriate technical and organizational measures to ensure and to be able to demonstrate that processing is performed in accordance with this Regulation. Those measures shall be reviewed and updated where necessary." |
| Article 25 | Data protection mechanisms should be embedded during the design stages of a system. A stated in the regulation "implement appropriate technical and organizational measures, such as pseudonymization, which are designed to implement data-protection principles, such as data minimization, in an effective manner and to integrate the necessary safeguards into the processing in order to meet the requirements of this Regulation and protect the rights of data subjects" In addition to that, the controller must ensure the protection of personal data (collected and analyzed) by default. |
| Article 26 | This article talks about the joint controller if needed in a system. The regulation says "Where two or more controllers jointly determine the purposes and means of processing, they shall be joint controllers. They shall in a transparent manner determine their respective responsibilities for compliance with the obligations under this Regulation," |
| Article 28 | This article specifies the data processor's responsibility. The processor needs to perform the task lawfully and legitimately with appropriate measures. One of the sections of Article 28-GDPR states "Where the processing is to be carried out on behalf of a controller, the controller shall use only processors providing sufficient guarantees to implement appropriate technical and organizational measures in such a manner that processing will meet the requirements of this Regulation and ensure the protection of the rights of the data subject" In addition, there are other regulations exists regarding the processing behavior and techniques of the processor |

## 4. Architectural Framework for COVID Vaccination Passport

### 4.1. Architectural Framework (VacciFi)

This section presents a proposal for a vaccination passport that is based on blockchain technology and inherits integrity by design. We have named the framework 'VacciFi'. Such a passport aims to allow the regional authorities to verify the vaccination status of a traveler. While considering the design requirements discussed earlier, the architectural framework presented in Figure 1 was formulated. The building blocks of the architectural framework include permissioned blockchain to allow certain actions to be performed only by certain identifiable actors. It also involves the use of smart contracts whenever some immigration authorities want to verify the vaccination details.

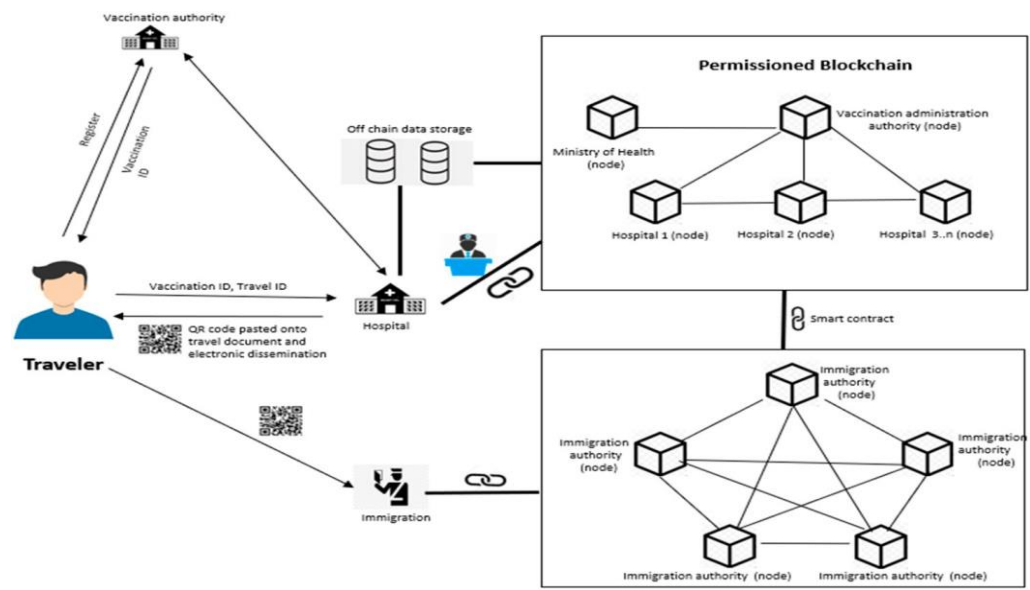

**Figure 1.** An architectural framework for VacciFi.

Briefly, the framework is based on the need that each citizen who intends to have a vaccination passport for traveling purposes has to register for the vaccine with their local health authority. Once the vaccine is administered, necessary details are recorded in the database (off-chain). Each potential traveler is given a QR code stamped on the passport and electronically via registered email for verification purposes. Whenever required, the QR code can be scanned for verifying vaccination details and validity. The process has 3 main steps discussed in the subsequent sub-sections.

Furthermore, one important aspect to discuss when it comes to data storage on the off-chain systems is access control management. For the VacciFi framework, we propose the following four operations for managing the access control:

- Create: refers to the right for the creation of a new record in the off-chain
- Read: refers to the right to read/view the record
- Update: refers to the right to update the existing record
- Delete: refers to deletion of the record

Based on these operations, different authorities will have different access privileges, which are detailed in Table 2.

**Table 2.** Access control privileges by different authorities.

|  | Create | Read | Update | Delete |
|---|---|---|---|---|
| **Immigration authority** |  | x |  |  |
| **Vaccination authority** | x | x |  | x |
| **Hospital authority** |  | x | x |  |
| **Ministry of health** |  | x |  |  |

As shown in the Table, the immigration authority has the right to read the existing records, however, the vaccination authority will have the ability to create, read and delete the records. The deletion capability is for off-chain storage, whenever such an operation is executed, it will perform logical deletion of the data stored on the blockchain as well. We discuss more details about this in Section 4.2.

4.1.1. Registration

To save precious lives, different countries have devised a vaccination policy consid- ering citizens more vulnerable to COVID. This policy mainly includes elder age groups, and vulnerable professionals such as healthcare workers and other frontline workers to be prioritized for vaccination. To get vaccinated the citizens have to be registered with the local health authority. This is the first step in the case of the proposed architectural framework. The registration can be done via different means, for example, online, using text messages, etc. After getting registered, the vaccination authority issues a vaccination ID, which has to be shared with the hospital staff to get vaccinated. When the user receives the vaccination ID, the consent conditions shall be included and encoded in a smart con- tract. The smart contract can also include other regulations, for example, the data retention period. The vaccination data can be kept as long as required by the local health authorities. The retention time needs to be included in the smart contract.

### 4.1.2. Vaccination

While visiting the healthcare facility, the citizens are required to carry identity docu- ments such as a passport and vaccination ID. The front desk would record all the necessary information about the person getting vaccinated. For GDPR compliance purposes, iden- tifiable information such as the traveler's name, passport number, contact details, etc. is stored off-chain. This will be supplemented by necessary access control mechanisms to ensure security. However, the data controller at the hospital stores two types of information onto the blockchain (1) the generated hash of the vaccination ID and passport number (2) the date of administering vaccination dose. As a result, a QR code is generated and pasted onto the passport and shared electronically with the traveler via email.

### 4.1.3. Verification

While traveling, the traveler presents the QR code to the immigration personnel. Scanning the QR code would retrieve all the relevant vaccination details along with the hash code generated during the vaccination phase from the local off-chain storage. The hash is then verified via smart contract by comparing with the one stored onto the permissioned blockchain. If there is a match, the blockchain returns dates of vaccination along with the validity. The details of the verification procedure are presented in Figure 2.

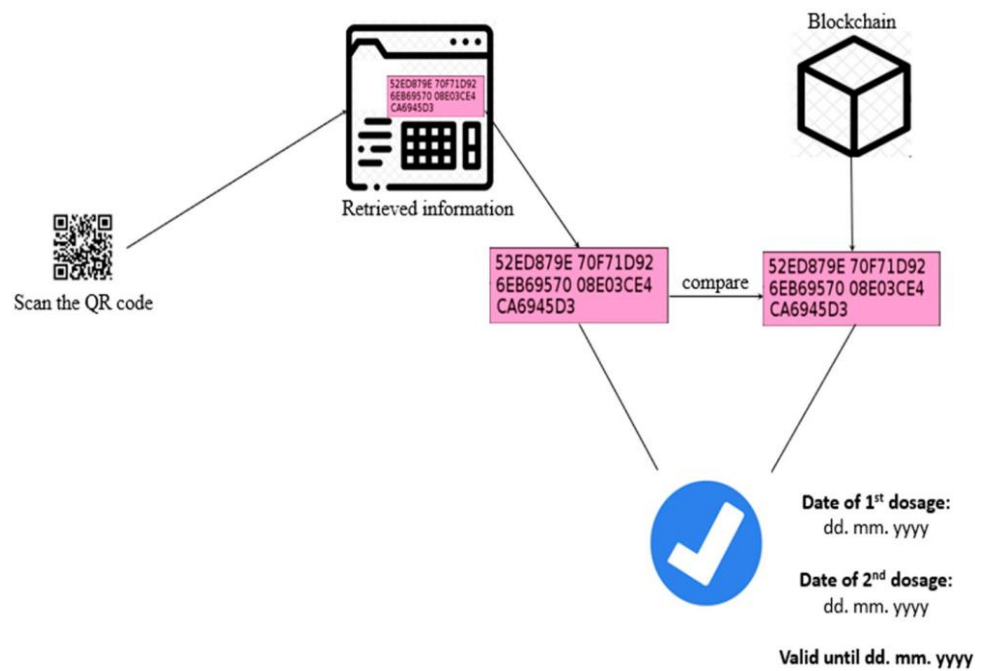

**Figure 2.** Description of the verification procedure.

### 4.2. Compliance with GDPR

The proposed architecture uses smart contracts for accessing vaccine information. The immigration authority and other concerned authorities will be able to retrieve the data by invoking smart contracts. From the GDPR compliance perspective, both the user and the hospital authorities will be informed about this event, which shall give compliance to Article 5 of GDPR. It is to note that the same process has also been recommended in prior literature [33].

GDPR specifically emphasizes the data subject's consent collection and its relationship with data analysis. This is crucial for the data controllers and processors. For compliance with GDPR, the use of smart contracts can be useful. The proposed architecture includes the consent conditions that will be designed and encoded into smart contracts. When the vaccine receiver registers with the vaccination authority the consent will be collected. This technique shall make our proposed system compliant with article 7 of GDPR [50,51].

Furthermore, the data deletion and modification procedure that is "Right to be for- gotten" and "Right to rectification" of GDPR Articles 17 and 16, respectively have to be considered. Whenever the vaccine receiver wants, he or she should be able to request the controller for data deletion as well as data correction in case any wrong information is entered. This problem can be solved by using off-chain storage for original data and on-chain storage for proof of existence [35–37,41]. A link will be established between the off-chain and on-chain storage. Whenever user requests for data deletion, this connection link will be destroyed causing logical deletion. It is worthwhile to mention that in the case of the proposed architecture, the off-chain data will also be deleted after logical deletion. So, there will be no personally identifiable information available on the off-chain. Besides, blockchain stores the hash only which is irreversible and is not understandable at all. So, it is impossible to retrieve or understand personal information only by looking at the hash.

Data controller and processor identification (Articles 24, 26, and 28) is another vital requirement for blockchain-based architectures. There are several recommendations for this purpose, for example, each node can act as a controller [37,38]. However, in some cases, miners with maximum financial stake can also act as controllers. If a smart contract is used in a system, smart contract developers can be also used as processors [39]. Since in the proposed architecture the permissioned blockchain is used, joint controllers (Article 26) could be used here. The miners who will be validating the blocks can be the processors (Article 28) [39,52,53]. Therefore, the proposed architecture is compliant with articles 26 and 28 of GDPR.

Some characteristics of blockchain are inherently compliant with the GDPR. For exam- ple, blockchain complies with the GDPRs privacy by design since it uses a cryptographic approach by default [54,55]. In addition, private blockchain helps implement security measures for cross-border data transmission.

## 5. Discussion

### 5.1. Research Contributions

This paper represents an architectural framework of GDPR-compliant vaccination passport that can be designed on a consortium blockchain. This architecture promotes a trustworthy and seamless process for vaccination status verification at border control or in any other required places.

Scalability is another crucial advantage of the proposed architecture. Since the ar- chitecture uses consortium blockchain, required authorities can be easily added to the network. The authorities can validate blocks using the Proof of Authority (POA) consen- sus algorithm.

Consortium blockchain consumes less computing power since the nodes are prede- fined. Moreover, the proposed solution uses POA, which also does not need a large amount of computational power. Hence, implementing the architecture will be economical in terms of computing power and installation cost. Besides other performance-related advantages, the POA consensus algorithm has advantages against 51% attack and distributed denial of service (DDoS) attacks. In POA consensus, the 51% attack requires the attacker to gain control over 51% of the network's nodes in contrast to others like proof of work consensus which requires control over 51% of the computation power. It is worth noting that obtaining control of the nodes in a permissioned blockchain network is much harder. The POA con- sensus algorithm defends against DDoS attacks since network nodes are pre-authenticated and block generation rights can be granted only to nodes having the ability to withstand the denial-of-service (DoS) attacks. In addition, in case of DoS attacks, the unavailable node can be removed from the network so that, it cannot take part in the validation process. Furthermore, the consortium blockchain helps alleviate anonymity. If there is any potential inside breach it can be detected very easily.

The architecture can implement customized data privacy policies as well as organi- zational policies if needed. The participating nodes can decide among themselves how the network will operate, and the validation process will be done. Furthermore, the train- ing process for using the network and validation takes less time since the number of stakeholders actively participating in the process is limited.

*5.2. Limitations*

This architecture is entirely a conceptual framework that can be developed with appropriate technical measures. Considering the stakeholder requirements in various organizations, the components inside the architecture and interaction among them can change. Moreover, real-time testing was not done, and the development phase is yet to be implemented.

One limitation of the proposed architecture is that it provides a mechanism for storing vaccination data, however, in line with the EU recommendations COVID passports should also be able to store negative COVID test results and recovery information from COVID in case of the previous infection.

Though the consortium blockchain uses pre-defined nodes for block validation, an in- sider attack is a possibility. If any node acts dishonestly it can hamper the validation process. One probable solution can be removing the dishonest nodes immediately upon finding out the malicious activity. Identifying the attacker is relatively easy in this architecture since the level of anonymity is much lower.

Another limitation is related to the GDPR's right to be forgotten policy. This is an issue related to the blockchain characteristics. Blockchain can act as an enabler of GDPR compliance, but there remains an argument that, as the data cannot be deleted from blockchain, it contradicts Article 17. We have addressed this issue here by storing only the hash of personal data onto the blockchain. Moreover, the architecture performs logical deletion and the transactions remaining on the blockchain cannot identify a data subject. In addition, after the logical deletion, original data is also deleted from the off-chain.

## 6. Conclusions

The paper presents a proposed architecture for a GDPR-compliant COVID vaccination passport (VacciFi). The architecture is scalable since the off-chain architecture is used for original data storage. The system shall provide availability, traceability, and integrity of vaccination information. VacciFi uses permissioned blockchain, which will enable a more controlled environment for the participating authorities. The data controller and processors are identifiable, so the data subjects can trust the system. To summarize, the proposed architecture can facilitate a secure, verifiable, and trustworthy environment that can be acclimatized with the current pandemic situation for ensuring safe travel and business across the border. Considering the limitations in the architecture, future work intends to (1) develop a proof of concept based on the proposed architecture; for this purpose we intend to collaborate with partner companies for development and testing as such; and (2) to scale up VacciFi for recording other COVID-specific details, as identified above and also for recording other vaccinations such as for polio, measles, etc. Such an implementation could be very helpful in tracking and verifying vaccinations worldwide post-pandemic.

**Author Contributions:** Conceptualization, A.B.H. and B.N.; methodology, A.B.H., B.N. and A.K.M.N.I.; validation, A.B.H., B.N., A.K.M.N.I. and S.H.; formal analysis, A.K.M.N.I. and S.H.; investigation, A.B.H., B.N.; data curation, A.K.M.N.I. and S.H.; writing—A.B.H. and B.N.; writing—review and editing, A.K.M.N.I. and S.H.; visualization, A.K.M.N.I. and S.H.; supervision, A.K.M.N.I. and S.H.; project administration, A.B.H., B.N., A.K.M.N.I. and S.H.; funding acquisition A.K.M.N.I. All authors have read and agreed to the published version of the manuscript.

**Funding:** This research was funded by BUSINESS FINLAND grant number 44559/31/2020 and The APC was funded by 44559/31/2020.

**Conflicts of Interest:** The authors declare no conflict of interest.